\documentclass[aps,prl,preprint,superscriptaddress]{revtex4-1}

\usepackage[utf8]{inputenc}
\usepackage[english]{babel}
\usepackage{amsmath}
\usepackage{amsfonts}
\usepackage{amssymb}
\usepackage{graphicx}
\usepackage{siunitx}
\usepackage[version=4]{mhchem}
\usepackage[normalem]{ulem}
\begin{document}

\title{A Unique Crystal Structure of Ca$_2$RuO$_4$ in the Current Stabilized Semi-Metallic State}

\author{J.~Bertinshaw}
\affiliation{Max Planck Institute for Solid State Research, Heisenbergstra{\ss}e~1, D-70569 Stuttgart, Germany}

\author{N.~Gurung}
\affiliation{Max Planck Institute for Solid State Research, Heisenbergstra{\ss}e~1, D-70569 Stuttgart, Germany}

\author{P.~Jorba}
\affiliation{Physik-Department, Technische Universit{\"a}t M{\"u}nchen, D-85748 Garching, Germany}

\author{H.~Liu}
\affiliation{Max Planck Institute for Solid State Research, Heisenbergstra{\ss}e~1, D-70569 Stuttgart, Germany}

\author{M.~Schmid}
\affiliation{Institute for Functional Matter and Quantum Technologies, University of Stuttgart, Pfaffenwaldring 57, D-70550 Stuttgart, Germany}
\affiliation{Center for Integrated Quantum Science and Technology, University of Stuttgart, Pfaffenwaldring 57, D-70550 Stuttgart, Germany}
\affiliation{Max Planck Institute for Solid State Research, Heisenbergstra{\ss}e~1, D-70569 Stuttgart, Germany}

\author{D.~T.~Mantadakis}
\affiliation{Max Planck Institute for Solid State Research, Heisenbergstra{\ss}e~1, D-70569 Stuttgart, Germany}

\author{M.~Daghofer}
\affiliation{Institute for Functional Matter and Quantum Technologies, University of Stuttgart, Pfaffenwaldring 57, D-70550 Stuttgart, Germany}
\affiliation{Center for Integrated Quantum Science and Technology, University of Stuttgart, Pfaffenwaldring 57, D-70550 Stuttgart, Germany}

\author{M.~Krautloher}
\affiliation{Max Planck Institute for Solid State Research, Heisenbergstra{\ss}e~1, D-70569 Stuttgart, Germany}

\author{A.~Jain}
\affiliation{Max Planck Institute for Solid State Research, Heisenbergstra{\ss}e~1, D-70569 Stuttgart, Germany}

\author{G.~H.~Ryu}
\affiliation{Max Planck Institute for Solid State Research, Heisenbergstra{\ss}e~1, D-70569 Stuttgart, Germany}

\author{O.~Fabelo}
\affiliation{Institut Laue Langevin, BP 156, F-38042 Grenoble cedex 9, France}

\author{P.~Hansmann}
\affiliation{Max Planck Institute for Chemical Physics of Solids, N{\"o}thnitzerstr Stra{\ss}e~40, D-01187 Dresden, Germany}
\affiliation{Max Planck Institute for Solid State Research, Heisenbergstra{\ss}e~1, D-70569 Stuttgart, Germany}

\author{G.~Khaliullin}
\affiliation{Max Planck Institute for Solid State Research, Heisenbergstra{\ss}e~1, D-70569 Stuttgart, Germany}

\author{C.~Pfleiderer}
\affiliation{Physik-Department, Technische Universit{\"a}t M{\"u}nchen, D-85748 Garching, Germany}

\author{B.~Keimer}
\affiliation{Max Planck Institute for Solid State Research, Heisenbergstra{\ss}e~1, D-70569 Stuttgart, Germany}

\author{B.~J.~Kim}
\affiliation{Max Planck Institute for Solid State Research, Heisenbergstra{\ss}e~1, D-70569 Stuttgart, Germany}
\affiliation{Department of Physics, Pohang University of Science and Technology, Pohang 790-784, South Korea}
\affiliation{Center for Artificial Low Dimensional Electronic Systems, Institute for Basic Science (IBS),
77 Cheongam-Ro, Pohang 790-784, South Korea}


\begin{abstract}
The electric-current stabilized semi-metallic state in the quasi-two-dimensional Mott insulator Ca$_2$RuO$_4$ exhibits an exceptionally strong diamagnetism. Through a comprehensive study using neutron and X-ray diffraction, we show that this non-equilibrium phase assumes a crystal structure distinct from those of equilibrium metallic phases realized in the ruthenates by chemical doping, high pressure and epitaxial strain, which in turn leads to a distinct electronic band structure. Dynamical mean field theory calculations based on the crystallographically refined atomic coordinates and realistic Coulomb repulsion parameters indicate a semi-metallic state with partially gapped Fermi surface. Our neutron diffraction data show that the non-equilibrium behavior is homogeneous, with antiferromagnetic long-range order completely suppressed. These results provide a new basis for theoretical work on the origin of the unusual non-equilibrium diamagnetism in Ca$_2$RuO$_4$.
\end{abstract}

\maketitle
The exploration of non-equilibrium phenomena in correlated-electron systems is a major frontier of condensed matter research. Most of the experimental work in this area has taken advantage of intense light fields, which were shown to induce non-equilibrium phase transitions in a wide variety of solids. Prominent examples include microwave-induced zero-resistance states in semiconductor quantum wells~\cite{Mani02}, Floquet states in topological insulators~\cite{Wang13}, and light-induced superconductivity in copper oxides~\cite{Faus11,Gedi07,Mank14}. Experiments on the antiferromagnetic (AFM) Mott insulator Ca$_2$RuO$_4$ recently unveiled a rare example of a phase transition induced by a DC voltage~\cite{Naka13,Okaz13,Sow17}. Under current flow, the insulating ground state was observed to transform into an electrically conducting phase with a high diamagnetic susceptibility~\cite{Sow17}.

The mechanisms responsible for this unusual insulator to metal transition (MIT) and the microscopic description of the non-equilibrium metallic phase are of intense current interest. Large diamagnetism can arise from light-mass Dirac electrons in semi-metals in the presence of strong spin-orbit coupling~\cite{Fuku70,Fuse15}, but such an electronic structure is incompatible with those of known metallic phases of ruthenates, which invariably have multiple large sheets of Fermi surface with four electrons evenly distributed over three orbitals, as is the case for the unconventional superconductor Sr$_2$RuO$_4$~\cite{Maen94,Carl12}. The transition to the insulating state by isovalent substitution of Sr for Ca involves a significant redistribution of electrons, which is reflected in a first order structural transition that involve the compression, tilt and rotation of the RuO$_6$ octahedra~\cite{Naka00,Frie01,Fang01}. This transition has also been identified in pressure~\cite{Naka02,Stef05,Tani13} and strain~\cite{Wang04,Diet18} studies of Ca$_2$RuO$_4$.

Thus, an important next step in our understanding of the current-induced state is an accurate knowledge of the atomic positions, which allows one to perform \textit{ab-initio} calculations of the electronic structure. In pump-probe experiments on light-induced phenomena, such information is difficult to obtain, although crystallographic studies have been reported in some cases~\cite{Gedi07,Mank14}. As the non-equilibrium phase in Ca$_2$RuO$_4$ is maintained in steady state by a DC current, it offers a rare opportunity to apply neutron crystallography, which is also a direct probe of magnetic structures. 

In this Letter, we use single crystal neutron diffraction to show that the non-equilibrium crystal structure of Ca$_2$RuO$_4$ is closely coupled to current density with behavior distinct from the first order transition that arises with other perturbation approaches. We also find that the AFM Bragg reflections are no longer present. Using the refined atomic positions of this state for density functional and dynamical mean field theories (DFT+DMFT) we find a high sensitivity of the electronic band structure to applied current --- for realistic values of the Coulomb interaction parameter~$U$ the resulting Fermi surface is partially gapped and includes small electron and hole pockets. This semi-metal band structure that potentially underlies the strong diamagnetism is not found in DMFT calculations of the equilibrium system, indicating the importance of the non-equilibrium crystal structure.

\begin{figure}
    \includegraphics[width=8.6cm,keepaspectratio]{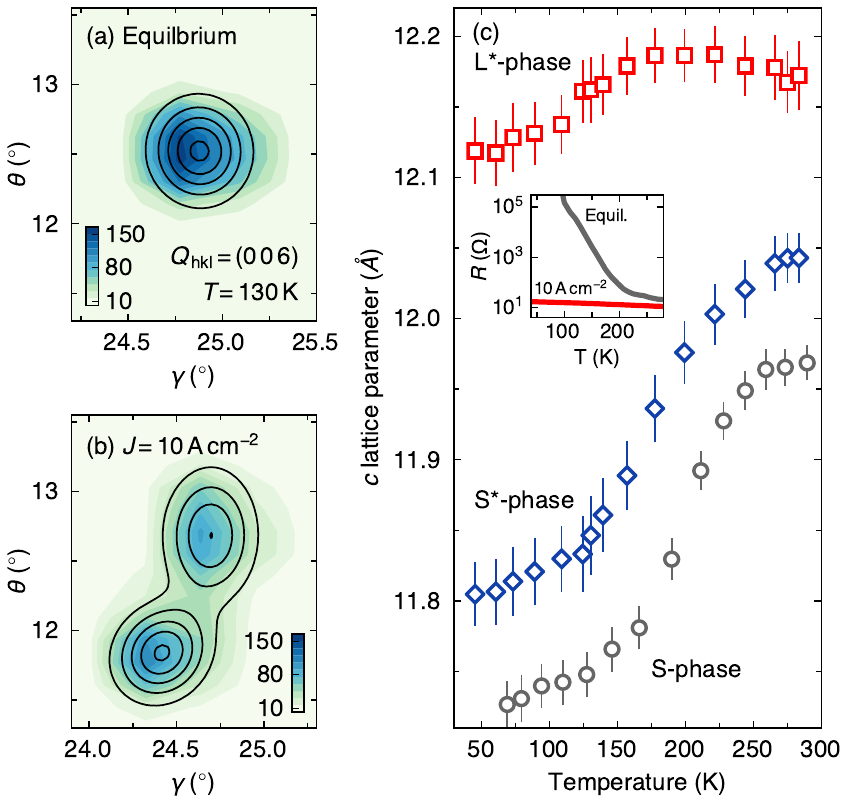}
    \caption{(a) Single crystal neutron diffraction rocking scan around the $(0\,0\,6)$ reflection at $T=130$\,K in the equilibrium state, shown as a map that plots the scattering angle $\theta$ versus the horizontal detector axis $\gamma$. (b) The same measurement under $J \parallel c\text{-axis} = 10$\,A\,cm$^{-2}$ reveals two separate peaks in the non-equilibrium state. (c) The fitted $Q$ position of the reflections, shown as contour lines in a) and b), was used to calculate the temperature trend of the $c$-axis lattice parameters. The non-equilibrium S*- and L*-phases display a behavior distinct from the equilibrium S-phase, shown in gray. The inset plots the \textit{in-situ} resistance of the two states.\label{FIG1}}
    \end{figure}

Neutron diffraction (ND) measurements were performed using instrument D9 at the Institut Laue-Langevin, Grenoble, France, with a high quality Ca$_2$RuO$_4$ crystal prepared using a floating zone mirror furnace with a RuO$_2$ self-flux in a process described previously~\cite{Naka01}. The untwinned crystal was mounted in such a way that \textit{in-situ} DC current was applied along the $c$-axis in a two-probe circuit utilizing a Keithley 2400 Source Measure Unit for sampling and control. A thermocouple was placed at the sample position and good thermal contact using silver epoxy was made in order to minimize the effects of Joule heating and ensure an accurate temperature reading.

At $T=280$\,K the applied voltage was systematically increased through the transition to the metallic regime, here defined as the characteristic step in the I-V curve shown in Supplemental Material~\footnote{See Supplemental Material for further information}. The current density was maintained at $J=10$\,A\,cm$^{-2}$ to ensure a steady state during temperature cycling (see the \textit{in-situ} resistance in the Fig.~\ref{FIG1}(c) inset). Rocking scans of the out-of-plane $(0\,0\,6)$ reflection were performed in $\sim$10\,K intervals from 290\,K through $T_N=110$\,K down to a lowest measurable temperature $T=45$\,K, limited by the cooling power of the cryostat that was competing with minor Joule heating of the sample and instrument wiring. The process was also conducted without applied current to compare with the equilibrium state.

The scans at $T=130$\,K are plotted in Figs.~\ref{FIG1}(a) and ~\ref{FIG1}(b) as colormaps that integrate over the vertical detector range (perpendicular to the scattering plane). The single reflection in equilibrium, corresponding to the stoichiometric `S-phase'~\cite{Naka97,Brad98}, splits into two reflections under $J=10$\,A\,cm$^{-2}$, indicating the presence of two phases. The integrated intensity revealed an almost equal volume fraction of the phases, which was fully restored after returning to the normal insulating state. The temperature evolution of the $c$-axis lattice parameter of the equilibrium and non-equilibrium phases, shown in Fig.~\ref{FIG1}(c), was calculated from a 2D Gaussian least-squares fitting of the $(0\,0\,6)$ reflections that are depicted as contour lines in Figs.~\ref{FIG1}(a,b). From the trends of the lattice parameters it is apparent that the phases in non-equilibrium are distinct from the equilibrium state, and are assigned here as the shorter $c$-axis S*-phase and the elongated L*-phase. 

The S*-phase $c$-axis expresses an expansion over the equilibrium S-phase that persists to the lowest measured temperature as it undergoes a significant contraction from 300\,K down to $\sim$150\,K. On the other side, the L*-phase undergoes a minimal contraction of the in the temperature region studied, which follows a trend similar to the equilibrium L-phase in previous perturbation studies~\cite{Frie01,Stef05}. It is important to point out that the nature of the S*-phase trend, notably the stabilization of the elongation below 150\,K, is not consistent with the effects of Joule heating and indicate that the current-stabilized system is not simply composed of an admixture of metallic and unaffected insulating regions.

\begin{figure}
    \includegraphics[width=8.6cm,keepaspectratio]{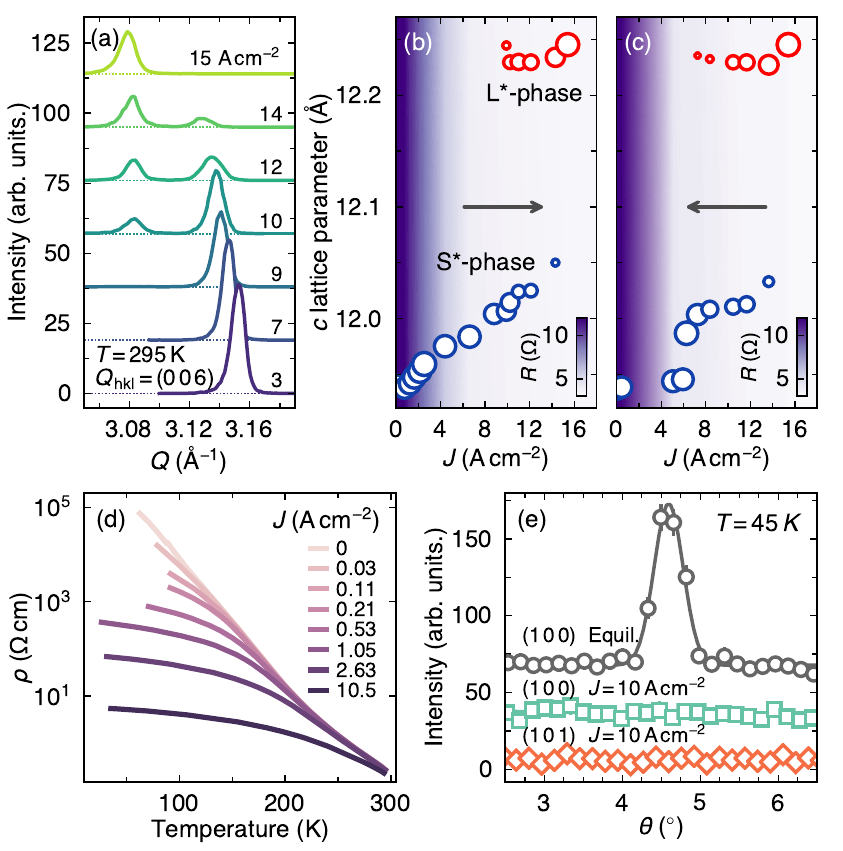}
    \caption{(a) XRD $2\theta$ scans of the $(0\,0\,6)$ reflection under increasing current density, applied along the $c$-axis, reveal the emergence of the L*-phase at higher densities (b) The $c$-axis lattice parameters of the S*-phase and L*-phase are plotted versus current density, with the symbol size indicating the integrated intensity and colormap representing the in-situ resistance. (c) A hysteresis in the structural behavior is seen upon reducing the applied voltage. (d) The 4-probe resistivity, with $V$$\perp$$c$, indicates a trend towards semi-metallic behavior under even minor current densities. (e) ND shows the $(1\,0\,0)$ magnetic reflection is suppressed at $T=45$\,K in the non-equilibrium state. The $(1\,0\,1)$ reflection that can arise due to an alternate AFM pattern is also not present.\label{FIG2}}
    \end{figure}

To investigate the current density dependence of the non-equilibrium phases, single crystal X-ray diffraction (XRD) and resistivity measurements were performed. XRD was conducted in ambient conditions using an in-house designed diffractometer with a Cu-K$\alpha$ source. The \textit{in-situ} voltage was applied using the same approach as the ND measurements. Figure~\ref{FIG2}(a) plots the momentum transfer $Q$ around the $(0\,0\,6)$ reflection, measured under increasing steps of current density $J$. The evolution of the out-of-plane $c$-axis lattice parameter was determined from the shift in the peak position, using a least-squares fit to a pseudo-Voigt function. The trend is shown in Fig.~\ref{FIG2}(b), along with the 2-probe sample resistance.

Starting from the equilibrium S-phase, the structure undergoes a continuous expansion of the $c$-axis with increasing applied current through the MIT. At larger current densities a second peak develops at smaller $Q$ around $J \approx 9.5$\,A\,cm$^{-2}$, indicative of the emergence of the L*-phase. As the current density increases further the overall scattered intensity transfers to the second phase, and at $J = 15$\,A\,cm$^{-2}$ the S*-phase phase is fully suppressed. A hysteresis in the structural behavior manifests when reducing the applied voltage, shown in Fig~\ref{FIG2}(c). Prior to the transition back to the insulating state, the S*-phase undergoes a distinct contraction, likely related to domain non-uniformity around the first-order transition as the L*-phase vanishes below $7$\,A\,cm$^{-2}$. Resistivity measurements were conducted using a 4-probe arrangement with $V$$\perp$$c$-axis in increasing steps of current density set at $T=290$\,K. The temperature curves, shown in Fig~\ref{FIG2}(d), reveal that Ca$_2$RuO$_4$ undergoes a continuous deviation away from the equilibrium Mott-insulating state with increasing current density. The XRD and resistivity measurements confirm previous reports by Nakamura~\textit{et.~al.}~\cite{Naka13} and Sow~\textit{et.~al.}~\cite{Sow17}, respectively. Combined with our ND study, these results show that the S*-phase evolves with current density and persists as the system becomes semi-metallic, before the L*-phase emerges and eventually dominates.

We corroborate these findings by showing that the non-equilibrium state studied with neutron diffraction does not exhibit the usual AFM structure found in equilibrium. Integer reflections were measured below and above $T_{\text{N}}$ at $T = 45$\,K and 130\,K, up to a momentum transfer $Q=4$\,\AA$^{-1}$, which revealed no sign of an AFM superstructure, notably the absence of the $(1\,0\,0)$ reflection associated with primary AFM order and the $(1\,0\,1)$ reflection that can arise due to an alternate AFM arrangement~\cite{Brad98}, as shown in Fig.~\ref{FIG2}(e). Additional Q-scans around these two reflections showed no sign of incommensurate order.

\begin{table}[b]
    \caption{\label{TAB1} Neutron diffraction structural refinement in the orthorhombic $Pbca$ space group. Ru--O bonds and RuO$_6$ octahedral parameters at $T=130$\,K of the S-phase and S*- and L*-phases at $J=10$\,A\,cm$^{-2}$. $\Theta$--O(1) refers to the tilt angle between the basal plane and the $ab$-plane, $\Theta$--O(2) is the angle between the Ru–-O(2) bond and the $c$-axis, and $\Phi$ is rotation of the RuO$_6$ around the $c$-axis. The Ru--O ratio compares the apical and averaged in-plane Ru-O bond lengths, and is a measure of the tetragonal distortion.}
    \begin{ruledtabular}
    \begin{tabular}{lccc}
    Phase  & S-phase & S*-phase & L*-phase \\
    Temperature (K)            & 130           & 130          & 130         \\
    \hline
    $a$~(\AA)                  &   5.3842(8)   &   5.404(4)   &   5.341(5)  \\
    $b$~(\AA)                  &   5.6158(9)   &   5.547(4)   &   5.436(6)  \\
    $c$~(\AA)                  &  11.7461(11)  &  11.848(8)   &  12.153(9)  \\
    Volume~(\AA$^3$)           & 355.16(3)     & 355.2(2)     & 352.8(3)    \\
    Orthorhombicity ($b$$-$$a$)&   0.23        &   0.14       &   0.10      \\
    \hline
    Ru--O(1)a~(\AA)            &   2.0132(11)  &   2.001(4)   &   1.964(4)  \\
    Ru--O(1)b~(\AA)            &   2.0161(10)  &   2.005(3)   &   1.968(5)  \\
    Ru--O(2)~(\AA)             &   1.9683(11)  &   1.979(4)   &   2.021(4)  \\
    Ru--O avg~(\AA)            &   1.999       &   1.995      &   1.984     \\
    Ru--O ratio                &   1.023       &   1.012      &   0.972     \\
    $\Theta$--O(1)~($^{\circ}$)&  12.79(1)     &  12.43(4)    &  10.69(4)   \\
    $\Theta$--O(2)~($^{\circ}$)&  11.53(1)     &  10.65(4)    &   9.76(4)   \\
    $\Phi$~($^{\circ}$)        &  11.965(3)    &  11.874(10)  &  12.034(11) \\
    \end{tabular}
    \end{ruledtabular}
    \end{table}

\begin{figure*}
    \includegraphics[width=14cm]{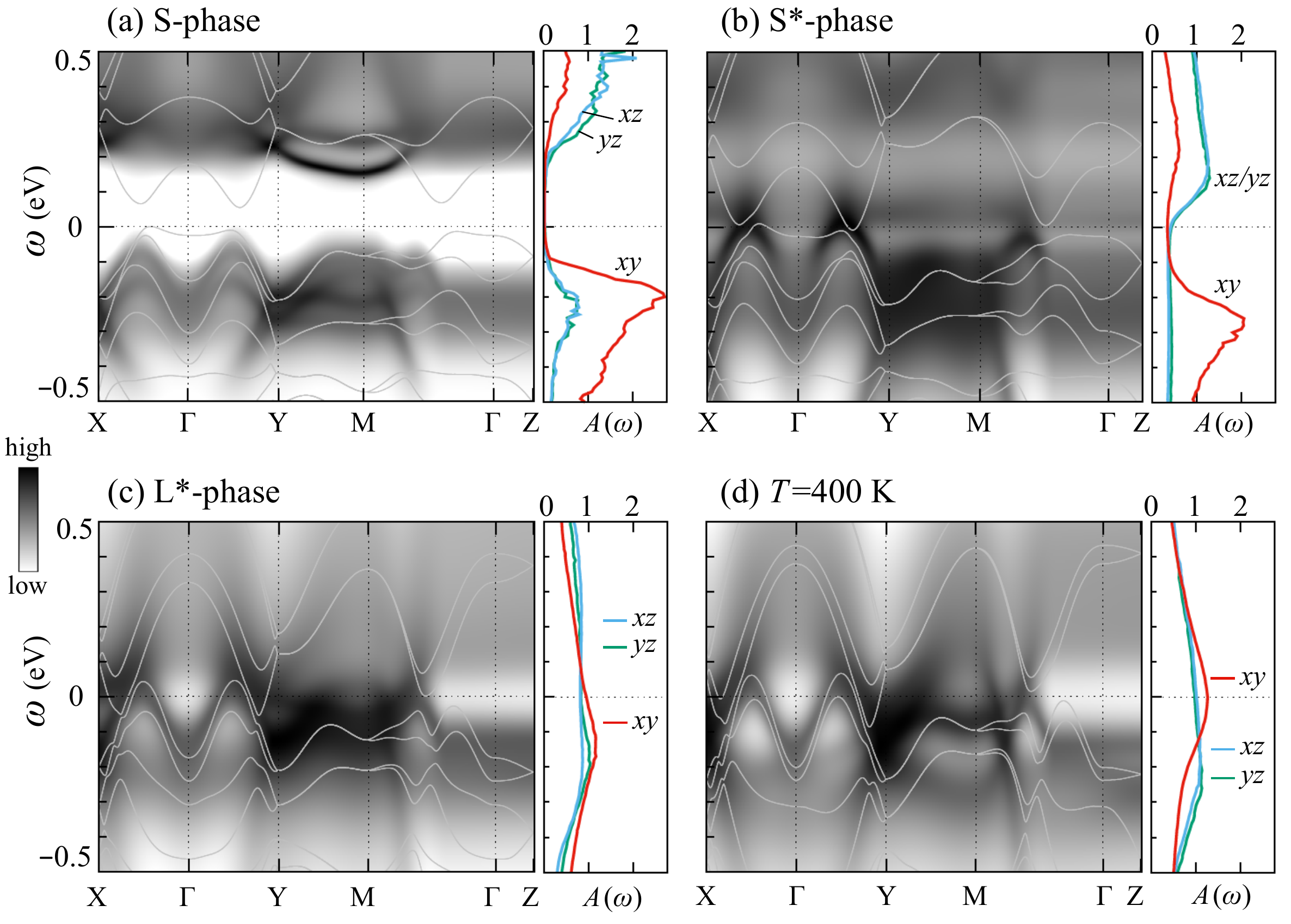}
    \caption{DMFT-calculated intensity map for electron spectral function $A(\omega,\mathbf{k})$ (in arbitrary units, dark color implies high intensity) as a function of energy $\omega$ (counted from chemical
    potential) and momentum $\mathbf{k}$ along high-symmetry directions in the orthorhombic Brillouin zone. Orbitally resolved local spectra $A(\omega)$ (in units of 1/eV) is shown on the right of each panel.
    Light gray lines show DFT+U mean-field bands. Calculations are based on the $T$\,=\,130\,K structure of the (a) equilibrium S-phase, non-equilibrium (b) S*- and (c) L*-phases, and on the (d) $T$\,=\,400\,K structure from Ref.~\onlinecite{Frie01}. In the current-induced phases (b) and (c), the insulating gap is closed and electron and hole pockets are formed, indicating semi-metallic behavior.\label{FIG3}}
    \end{figure*}

A precise determination of the crystallographic details of the S*- and L*-phases was conducted using an extensive range of reflections collected with ND at $T=130$\,K and $J=10$\,A\,cm$^{-2}$. Noting that no additional magnetic or nuclear reflections were identified at both 45\,K and 130\,K under applied current, least squares refinement was conducted using using the $Pbca$ space group across the $T=130$\,K equilibrium and non-equilibrium phases. Sets of $\sim$300 reflections were used for the refinement of each phase individually, using the \textsc{fullprof} software suite. The primary results are shown in Table~\ref{TAB1}, with atomic positions and fit quality listed in Supplemental Material~\cite{Note1}.

In the S-phase, tetragonal and orthorhombic distortions have three primary effects upon the RuO$_6$ octahedra---a $c$-axis flattening, tilting of the basal plane through the $ab$-plane and a rotation around the $c$-axis~\cite{Brad98}. The S-phase Mott-state and establishment of AFM are closely coupled to the degree of the orthorhombicity and tetragonal compression~\cite{Fang01,Jain17}. Indeed, through perturbation, such as temperature or Sr-ion substitution, the equilibrium system undergoes a first order transition to a metallic L-phase, where the $c$-axis elongates dramatically. At the same time, the overall unit cell volume decreases as the $a,b$-axis lattice parameters tend towards parity, leading to octahedral elongation and reduced tilt and rotation angles (with Sr$_2$RuO$_4$ the tetragonal and undistorted end member)~\cite{Frie01,Stef05}. The non-equilibrium L*-phase shares behavior with these equilibrium L-phases, including a first-order transition behavior and a reversal of the ratio between the basal and apical bond lengths.  It is notable then that the non-equilibrium S*-phase atomic positions reveal a marked decrease in the orthorhombicity and reduced tetragonal compression, even as the $c$-axis lattice parameter expansion is minor in this regime (see Fig.~\ref{FIG1} and Ref.~\onlinecite{Naka13}), displaying behavior distinct from other perturbation approaches.

To study the sensitivity of the electronic state to the crystallographic distortions in the equilibrium and non-equilibrium phases, we have conducted band structure calculations utilizing the refined crystal structures and including spin-orbit coupling effects. Electron correlations are treated on DMFT level, using representative values of Coulomb repulsion $U$\,=\,1.9\,eV and Hund's coupling $J_H$\,=\,0.4\,eV~\cite{Sutt17}. Figure 3 shows the calculated spectral function $A(\omega,\mathbf{k})$ and its $\mathbf{k}$-integrated value $A(\omega)$ (i.e., local density of states) near the Fermi level. The results for a wider energy interval, and full computational details can be found in Supplementary Material~\cite{Note1}. We note that these calculations are in good agreement with experimental ARPES data~\cite{Sutt17,Ricc18}. For comparison, we have also plotted DFT+U mean-field band dispersions.

A current-driven insulator-to-metal transition is revealed by our calculations. The equilibrium S-phase is found to be Mott-insulating, with a gap $\sim$\,0.2\,eV between lower (upper) Hubbard bands of predominantly $d_{xy}$ ($d_{xz/yz}$) orbital character, see $A(\omega)$ in Fig.~\ref{FIG3}(a). In the non-equilibrium (b) S*- and (c) L*-phases, these bands broaden and overlap, releasing hole and electron charge carriers. The overlap is very small in the S*-phase (b), suggesting a semi-metallic state with the hole and electron pockets, derived from $d_{xy}$- and $d_{xz/yz}$-orbitals correspondingly. We note, however, that spin-orbit coupling and low-symmetry distortions somewhat mix the orbital content of these pockets. 

The electronic states in the L*-phase (c), where tetragonal and orthorhombic distortions are further reduced, resemble those of $T$\,=\,400\,K structure (d). In particular, density of states near the Fermi level has no pseudogap feature and all $t_{2g}$ orbitals contribute nearly equally, see $A(\omega)$ in Fig.~\ref{FIG3}(c) and (d), as expected in a metallic state with small orbital disproportionation. We find the average orbital occupations of $n_{xy}=1.47 (1.34)$, $n_{xz}=1.28 (1.34)$, and $n_{yz}=1.25 (1.32)$ for the L*-phase ($T$\,=\,400\,K structure).

These results reveal that the electronic band structure is extremely sensitive to current density through the RuO$_6$ distortions, making it clear that the structural details must be considered in any model to describe the mechanism that drives the anomalous diamagnetism under direct current, such as the Dirac point formation proposed by Sow~\emph{et. al.}~\cite{Sow17}.

In summary, we used neutron crystallography to determine the structure of non-equilibrium Ca$_2$RuO$_4$ and identify a phase associated with unique magnetic, structural and electronic properties. Our experimentally determined atomic coordinates and the electronic structure in the semi-metallic state indicated by the DMFT calculations provide a new basis for theoretical work on the origin of the unusual non-equilibrium diamagnetism of Ca$_2$RuO$_4$. The conspicuous deviation from the usual trend of the metallic L-phases found in equilibrium indicates a unique mechanism for the current-induced state, and opens a new experimental approach to tune Ca$_2$RuO$_4$ and to understand the competing interactions underlying $4d$ TMOs.

We would like to thank J.~Porras, C.~Sow and Y.~Maeno for fruitful discussions. We acknowledge financial support by the European Research Council under Advanced Grant No. 669550 (Com4Com). J.B. was supported by the Alexander von Humboldt Foundation.

%

\end{document}